\documentclass[%
 reprint,
 amsmath,amssymb,
 aps,
]{revtex4-2}

\usepackage{amsmath}
\usepackage{graphicx}
\usepackage{dcolumn}
\usepackage{bm}
\usepackage[utf8]{inputenc}
\usepackage[T1]{fontenc}
\usepackage{mathptmx}
\usepackage[T1]{fontenc} 
\usepackage{amsmath,color}
\usepackage{amssymb}
\usepackage{amsfonts}
\usepackage{amsthm}
\usepackage{mathtools}
\usepackage[title]{appendix}
\usepackage{hyperref}
\usepackage{braket}
\usepackage{xcolor}
\usepackage{etoolbox}
\usepackage{hyperref}
\usepackage{xfrac}
\usepackage{soul}
\usepackage{mathrsfs}
\usepackage[mathcal]{eucal}
\usepackage{tabularx,booktabs,array}
\newcolumntype{Y}{>{\centering\arraybackslash}X}

\usepackage{algorithm}
\usepackage{algpseudocode}
\usepackage{dsfont}

\usepackage{bbold}

\begin{document}
\preprint{AIP/123-QED}

\title{A Resolution to the Marcus–Rehm–Weller Paradox in Electron Transfer}
\author{Ethan Abraham}
\email{ethana@mit.edu}
\affiliation{Department of Chemistry, Massachusetts Institute of Technology, Cambridge, Massachusetts 02139, USA}\

\date{First posted 31 October 2025; revised 25 November 2026}

\begin{abstract}
Marcus theory famously predicts that electron-transfer rates decrease once the thermodynamic driving force exceeds the reorganization energy. Yet many systems instead exhibit Rehm–Weller kinetics, characterized by rate saturation rather than decrease. Here we show that these apparently contradictory phenomenologies can emerge as opposite physical limits of the same two-state harmonic Hamiltonian. In the normal region, the model recovers both Marcus and Rehm–Weller behavior. In the inverted region, however, it predicts Marcus’s decreasing rate in the nonadiabatic limit but Rehm–Weller saturation in the adiabatic limit. Using physically plausible reorganization energies and electronic coupling values, we show that Rehm–Weller’s data can be quantitatively reproduced within a microscopic quantum model. Finally, we note that the model predicts three regions of adiabatic electron transfer: (1) a normal region with kinetics governed by an effective reorganization energy, (2) a finite region of barrierless transfer, and (3) a region in which the correspondence between diabatic and adiabatic states is reversed, suppressing the Marcus rate.
\end{abstract}

\maketitle

While electrical conductivity in metals involves electron flow via delocalized states within partially filled bands \cite{ashcroft_mermin}, many important processes involve electron-transfer (ET) between discrete molecular orbitals, such as ET in batteries \cite{CIET_LCO,LFP_images}, electrocatalysis \cite{Qin2023_CO2_Reduction,zhang_driving_2020}, solar energy conversion \cite{Bredas_Review}, and biological redox \cite{Marcus_1985}. 

The physics of nonadiabatic ET was outlined by Marcus in his seminal 1956 paper and subsequent works \cite{Marcus_1956,Marcus_1964,Marcus_1965}. In modern language, Marcus theory describes a weakly coupled two-state electronic system interacting with a slow collective nuclear coordinate representing polarization response to the charge localization. In the nonadiabatic limit, the small donor–acceptor coupling permits a separation of nuclear and electronic motion into sequential steps: (1) thermal fluctuation of the nuclear coordinate to the diabatic crossing point, (2) an effectively instantaneous Franck–Condon electronic transition at fixed nuclear positions, and (3) relaxation on the product free-energy surface. Approximating the diabatic free energies as harmonic functions of the collective solvent coordinate leads to the iconic expression for the activation barrier \cite{Marcus_1956,Marcus_1964,Marcus_1965,Nitzan_2006}.

Famously, the derived rate expression predicts that although the ET rate at first (expectedly) increases as the thermodynamic driving force is increased, once a critical driving force is reached, the rate decreases, a phenomenon known as the Marcus inverted region \cite{marcus1960theory,Marcus_1965,Marcus_1964}. However, early experiments often exhibited a saturation of the rate at large driving force rather than this predicted decrease, as notably discussed by Rehm and Weller \cite{experiment_Weller} and others \cite{experimental_Foote,experiment_Scheerer}. In 1984, the inverted phenomenon was first demonstrated experimentally by Miller \textit{et al.} \cite{Miller_inverted}, and, after many subsequent demonstrations \cite{Wasielewski_inverted,experimental_Giovanny,experimental_Sun}, Marcus's picture has been accepted as standard \cite{Nobel_Marcus} while the experiments exhibiting the Rehm–Weller phenomenon are conventionally rationalized as reaching a diffusion-limit prior to the onset of the Marcus inverted region \cite{Rosspeintner_diffusion,Vauthey_2020,Photo_textbook}. However, this explanation is not fully satisfying, for the ET rate should be a composition of both a diffusion-controlled encounter frequency and an Arrhenius activation factor, and the latter is often unaddressed. Several additional explanations for the absence of the inverted region have also been proposed, including alternative ET pathways towards an ionic product in an excited state (i.e. when ET towards the ground state would be in the inverted region) \cite{experiment_Weller}, as well as breakdowns of the linear dielectric response approximation \cite{yoshimori1989shapes}, distance dependence of solvent reorganization energies \cite{tachiya1992new,burshtein2007diffusional,rosspeintner2008rehm,Rosspeintner_diffusion}, or instabilities of reaction–diffusion in reactive mixtures with negative differential resistance \cite{bazant2017thermodynamic}.

Here, we propose another possible explanation for the commonly observed absence of the inverted region, which is that if one interprets the Marcus parabolas as forming a harmonic two-state quantum system (as assumed in many treatments \cite{Nobel_Marcus,Nitzan_2006,Bredas_Review,Brunschwig_Book,Fay_extending}), the same physical model that predicts Marcus kinetics in the nonadiabatic limit also predicts Rehm–Weller kinetics in the adiabatic limit. 

We begin by briefly reviewing a standard mathematical treatment of the harmonic two-state quantum system that has been described in many references \cite{Ethan_lambda_eff,Ratner_review,Marcus_1985,Brunschwig_Book,Bredas_Review,Fay_extending,Nitzan_2006}. The system is composed of donor and acceptor diabatic electronic states $\ket{\phi_D}$ and $\ket{\phi_A}$ with energies given by \begin{subequations}\label{energies}\begin{equation}\label{ED} E_D(q)=\lambda q^2,\end{equation}\begin{equation}\label{EA}E_A(q)=\lambda(1-q)^2+\Delta E,\end{equation}\end{subequations} where $\lambda $ is the diabatic reorganization energy defined in Marcus theory, $q$ is a collective nuclear coordinate defined to satisfy $\langle q\rangle_D=0, \langle q\rangle_A=1$, and $\Delta E$ is the driving force. Setting the two energies equal, we obtain the coordinate and energy of the transition state (TS,*) in the nonadiabatic (i.e. $V\rightarrow0$) limit, \begin{equation}\label{marcus_ts}q^*=\frac{1}{2}\left(1+\frac{\Delta E}{\lambda}\right),\end{equation}\begin{equation}\label{marcus_Ea}E^*(\Delta E)=\frac{(\lambda+\Delta E)^2}{4\lambda},\end{equation} a central result of Marcus theory \cite{Marcus_1956,BixonJortner1999,Bredas_Review,Ratner_review,Nitzan_2006}. This expression famously predicts a decrease in the rate as the driving force is increased beyond $-\lambda,$ as has been observed experimentally \cite{Miller_inverted,Wasielewski_inverted,experimental_Giovanny,experimental_Sun}.

Beyond the nonadiabatic limit, the diabatic states can mix via an electronic coupling $V(q)$. Here we take constant $V(q)=V$ for simplicity, as commonly known as the Condon approximation \cite{Marcus_1985,Marcus_1989,Nitzan_2006,Blumberger_Condon,Troy_Condon}. We can then cast Eq. (\ref{energies}) as the two-state Hamiltonian in the diabatic basis $\{\ket{\phi_D},\ket{\phi_A}\}$ with the off-diagonal element defined by the electronic coupling \cite{Nitzan_2006,Blumberger_Condon,Jianshu_spectral,Fay_extending,VanVoorhis_2010,Newton_probes,Hush_1997,Hush_1999,Ethan_lambda_eff}\begin{equation} \label{matrix}H(q)=
\begin{pmatrix}
E_D(q) & V \\
V & E_A(q)
\end{pmatrix}=\begin{pmatrix}
\lambda q^2 & V \\
V & \lambda(1-q)^2+\Delta E
\end{pmatrix}.
\end{equation} Solving the secular equation gives the energies of the corresponding states in the adiabatic basis $\{\ket{\psi_-
},\ket{\psi_+}\}$ \begin{equation}\label{adiabats}\begin{split}E_\pm(q)&=\frac{E_D(q)+E_A(q)}{2}\pm\frac{1}{2}\sqrt{(E_D(q)-E_A(q))^2+4V^2}\\&=\frac{\lambda(2q^2-2q+1)+\Delta E}{2}\pm\\&~~~~~~~~~~~~~~~~~~~~~~~~~\frac{1}{2}\sqrt{(\lambda(2q-1)-\Delta E)^2+4V^2},\end{split}\end{equation} from which the exact activation barrier can be obtained \begin{equation}\label{Ea_exact}E^*=E_-(q^*)-E_-(q_\text{r}),\end{equation} where $q^*$ is the coordinate of the TS and $q_\text{r}\approx0$ the coordinate of the reactant minimum \cite{Ethan_lambda_eff,Nitzan_2006}.

\begin{figure}[t]
\includegraphics[width=1\columnwidth]{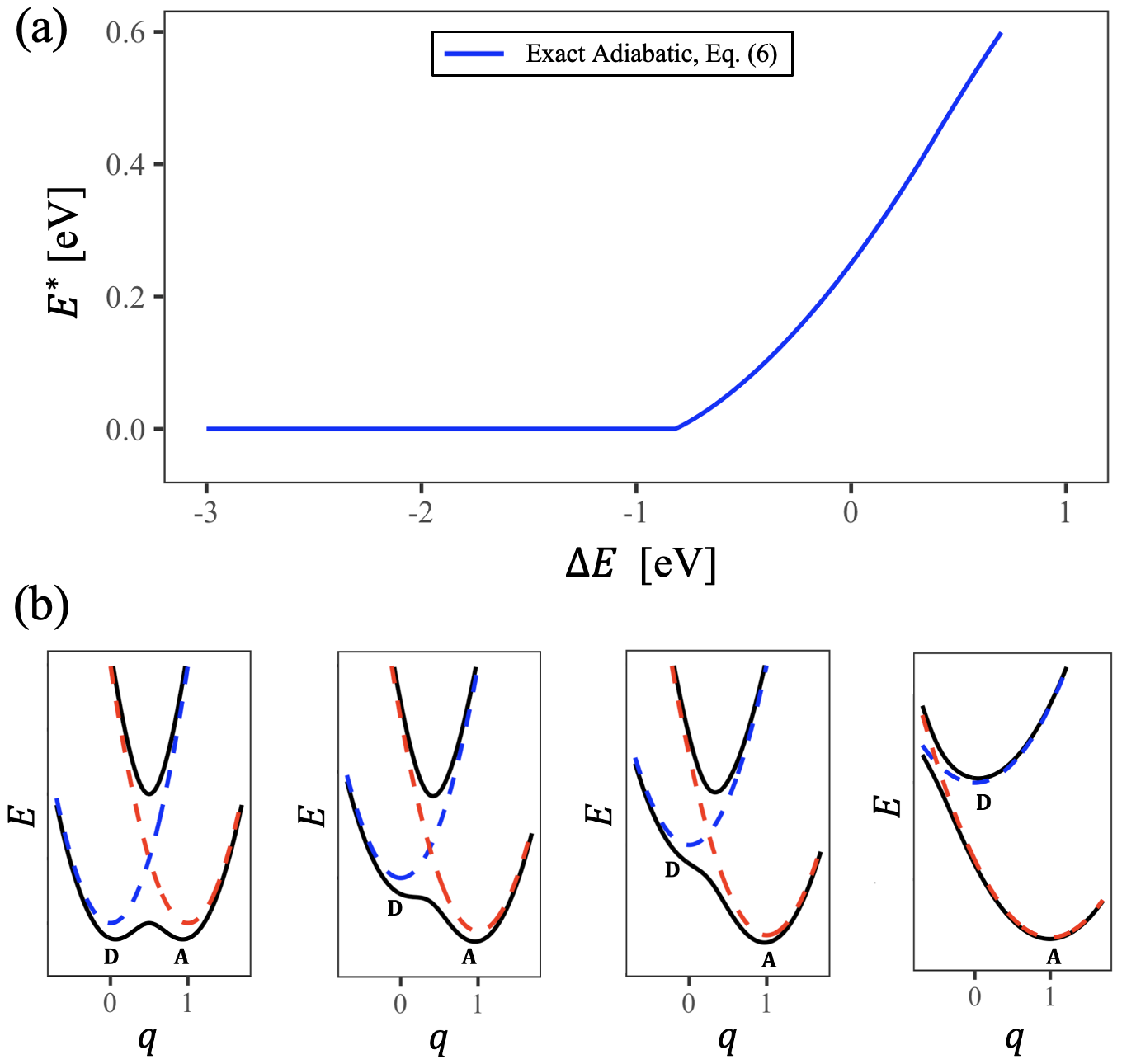}
\caption{(a) The activation barrier calculated from Eq. (\ref{Ea_exact}) plotted against the driving force. Parameter values are set to $\lambda=4$ eV and $V=1$ eV. (b) Schematics of the energy of the adiabats $E_\pm(q)$ (solid black) and the donor (blue dashed) and acceptor (red dashed) diabats plotted as a function of the reaction coordinate $q$. The panels from left to right correspond to increasing driving force, ranging from $\Delta E=0$ (left) to $\Delta E<-\lambda$ (right).}
\label{FIG0}
\end{figure}

While the nonadiabatic activation barrier is frequently calculated as a function of the driving force using Eq. (\ref{marcus_Ea}), the exact adiabatic activation barrier can be calculated numerically using Eq. (\ref{Ea_exact}). In Fig. \ref{FIG0}(a) this calculation has been performed for our model system with parameter values set to $\lambda=4$ eV and $V=1$ eV, with those specific parameters chosen only for illustrative purposes (the large value of $V$ is chosen to represent a strongly adiabatic reaction). We observe a dependence of the activation barrier on the driving force consistent with Rehm–Weller kinetics. In particular, the activation barrier decreases gradually up to a point and then flattens into a plateau of barrierless transition, consistent with the observations in the Rehm–Weller experiment \cite{experiment_Weller}. Figure \ref{FIG0}(b) illustrates this behavior by drawing in the adiabats corresponding to the diabatic Marcus parabolas, where we observe a range of driving forces corresponding to barrierless transitions after a certain driving force is reached (see the middle two panels). So our proposed resolution to the Marcus–Rehm–Weller paradox is a simple one: If the reaction has strong electronic coupling, Marcus's model actually predicts Rehm–Weller kinetics. 

However, our task is not yet complete. The careful reader will notice that this resolution to the paradox leads to a follow-up question: If the reaction is adiabatic, why are Marcus kinetics observed in the normal region?

\begin{figure}[!b]
\includegraphics[width=1\columnwidth]{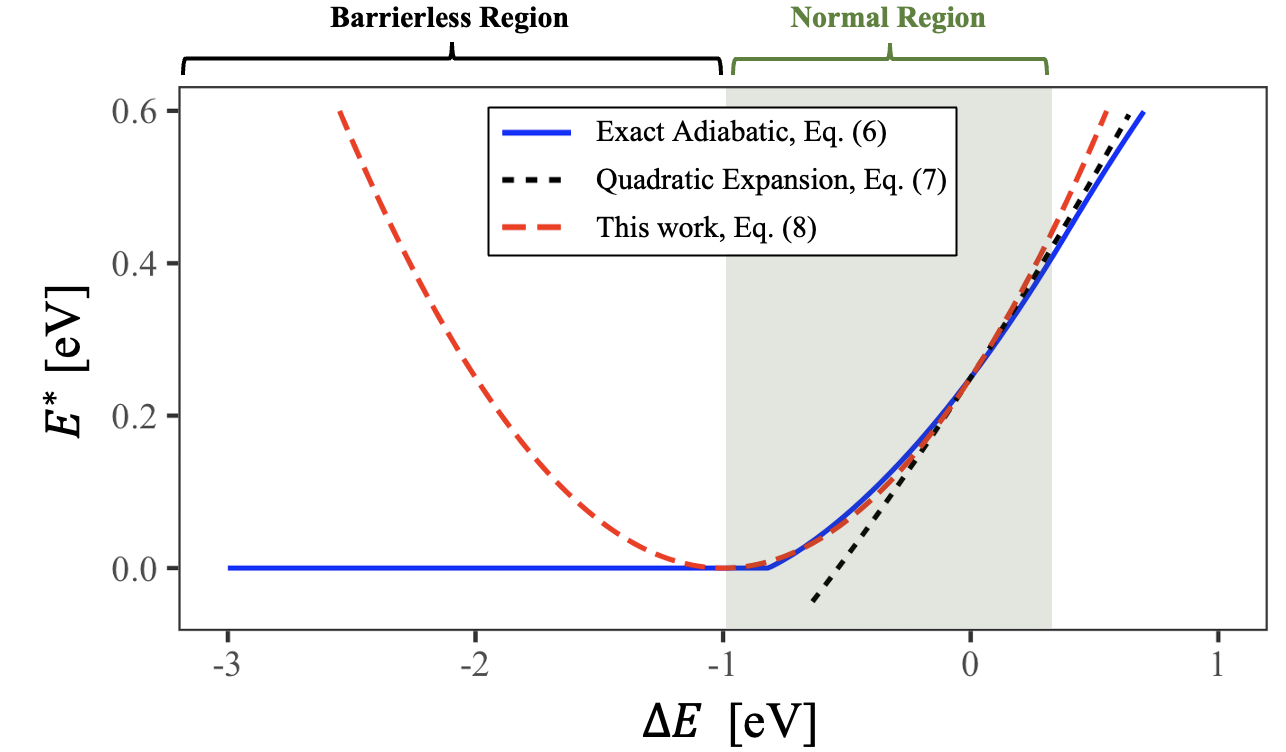}
\caption{The solid blue line is the same as in Fig. \ref{FIG0}, corresponding to the activation barrier calculated according to Eq. (\ref{Ea_exact}) as a function of the driving force. The dashed black line shows the corresponding activation barrier calculated from Eq. (\ref{ts}), truncated at quadratic order. The red dashed line shows the corresponding activation barrier calculated from Eq. (\ref{Discovery}), obtained from multiplying $(\Delta E)^2/4\lambda$ in Eq. (\ref{ts}) by the series $\sum_{n=1}^\infty n(2V/\lambda)^{n-1}$. This simple transformation leads to close agreement between the resulting polynomial and the exact adiabatic barrier throughout the normal region, which is shaded in the figure.}
\label{FIG01}
\end{figure}

This second question can be resolved via a remarkable mathematical property of the Marcus system, which is that the activation barrier of Eq. (\ref{Ea_exact}) exhibits a Marcus-like dependence on the driving force throughout the full normal region even for strongly coupled reactions \cite{Ethan_lambda_eff}.

To demonstrate this mathematical fact, we begin with the quadratic expansion of Eq. (\ref{Ea_exact}) following previous references \cite{Ethan_lambda_eff,Blumberger_pert,Brunschwig_Book} \begin{equation}\label{ts}\begin{split} E^*=\frac{\lambda}{4}+\frac{\Delta E}{2}&+\frac{(\Delta E)^2}{4\lambda}-V+\frac{V^2}{\lambda} +\\&O\left( \frac{V^3}{\lambda^2}, \frac{V^2 \Delta E}{\lambda^2}, \frac{V (\Delta E)^2}{\lambda^2}, \frac{(\Delta E)^3}{\lambda^2} \right). \end{split}\end{equation}
The dashed black line in Fig. \ref{FIG01} shows the activation barrier from this quadratic expansion, omitting all higher order terms, as a function of the driving force. It is overlaid atop the solid blue line from Fig. \ref{FIG0}. We notice that the agreement of this quadratic expansion with the exact adiabatic barrier is good in the vicinity of zero driving force, but it is rather poor as the plateau is approached. 

Given the poor accuracy of the quadratic expansion as the plateau is approached, it is natural to consider what higher-order terms would improve the agreement. Remarkably, if one multiplies the $(\Delta E)^2/4\lambda$ term in the quadratic expansion by the series $\sum_{n=1}^\infty n(2V/\lambda)^{n-1}$, the dashed red curve in Fig. \ref{FIG01} is obtained. This curve closely follows the exact adiabatic barrier throughout the normal region. The equation for this curve is \begin{equation}\label{Discovery}\begin{split} E^*=\frac{\lambda}{4}+\frac{\Delta E}{2}&+\frac{(\Delta E)^2}{4\lambda}\left[\sum_{n=1}^\infty n\left(\frac{2V}{\lambda}\right)^{n-1}\right]-V+\frac{V^2}{\lambda}. \end{split}\end{equation}

Equation (\ref{Discovery}) can now be simplified into a suggestive form. We obtain 
\begin{equation}\label{compact}\begin{split} E^*&=\frac{\lambda}{4}+\frac{\Delta E}{2}+\frac{(\Delta E)^2}{4\lambda(1-2V/\lambda)^2}-V+\frac{V^2}{\lambda}\\&=\frac{\lambda(1-2V/\lambda)^2}{4}+\frac{\Delta E}{2}+\frac{(\Delta E)^2}{4\lambda(1-2V/\lambda)^2}. \end{split}\end{equation} Defining $\Lambda=\lambda(1-2V/\lambda)^2$, Eq. (\ref{compact}) folds into the familiar driving force dependence \cite{Ethan_lambda_eff,Marcus_1985}
\begin{equation}\label{result} E^*(\Delta E)=\frac{(\Lambda+\Delta E)^2}{4\Lambda}.\end{equation} We refer to $\Lambda$ as the \textit{effective reorganization energy}, first introduced in Ref. \cite{Ethan_lambda_eff}, because it takes the place of the familiar Marcus reorganization energy in Eq. (\ref{result}) and approaches the Marcus reorganization energy in the nonadiabatic limit. Questions regarding the dynamical meaning and/or philosophical interpretation of this parameter are beyond the scope of this work; for now the point is only that we have identified a parameter that accurately interpolates between Marcus and Rehm–Weller kinetics. 

We note that the infinite series in Eq. (\ref{Discovery}) is motivated by our ansatz that a Marcus-like driving-force dependence holds beyond the nonadiabatic limit \footnote{Our central equation has the three properties that (i) it retains the Marcus-like functional form, (ii) it agrees with Eq. (6) at $V=0$ for all $\Delta E$, and (iii) it agrees with Eq. (6) at $\Delta E=0$ for all permitted $V<\lambda/2$.}. That such a simple construction tracks the adiabatic barrier across the entire normal region is a mathematical observation, one on which the theoretical significance of this work rests. The construction requires the convergence criterion $|2V/\lambda|<1,$ which the parameters used in this work satisfy. Notably, this is also the criterion under which the lower adiabat is double-welled at zero driving force, since for $2V\geq\lambda$ the surface has a single minimum \cite{Jianshu_spectral,Brunschwig_Book}.

Comparing Eq. (\ref{result}) with Fig. \ref{FIG01}, we obtain a simple piecewise rate expression for Rehm–Weller kinetics

\begin{equation}\label{Rehm–Weller}k_{\text{RW}}(\Delta E)=
\begin{cases}
C\text{exp}\left[ -\frac{(\Lambda+\Delta E)^2}{4 k_BT\Lambda} \right], & \Delta E \ge -\Lambda \\
~~~~~~~~~~~~C,& -\lambda < \Delta E < -\Lambda
\end{cases}\end{equation} where $C$ is a driving-force independent prefactor frequently associated with the diffusion-controlled encounter limit \cite{Thermodynamic_and_Kinetic_Limits,Vauthey_2020}. The third region $\Delta E < -\lambda$, where the rate is suppressed, is discussed later with Eqs.~(\ref{inverted:13})--(\ref{inverted:16}).

\begin{figure}[b]
\includegraphics[width=1\columnwidth]{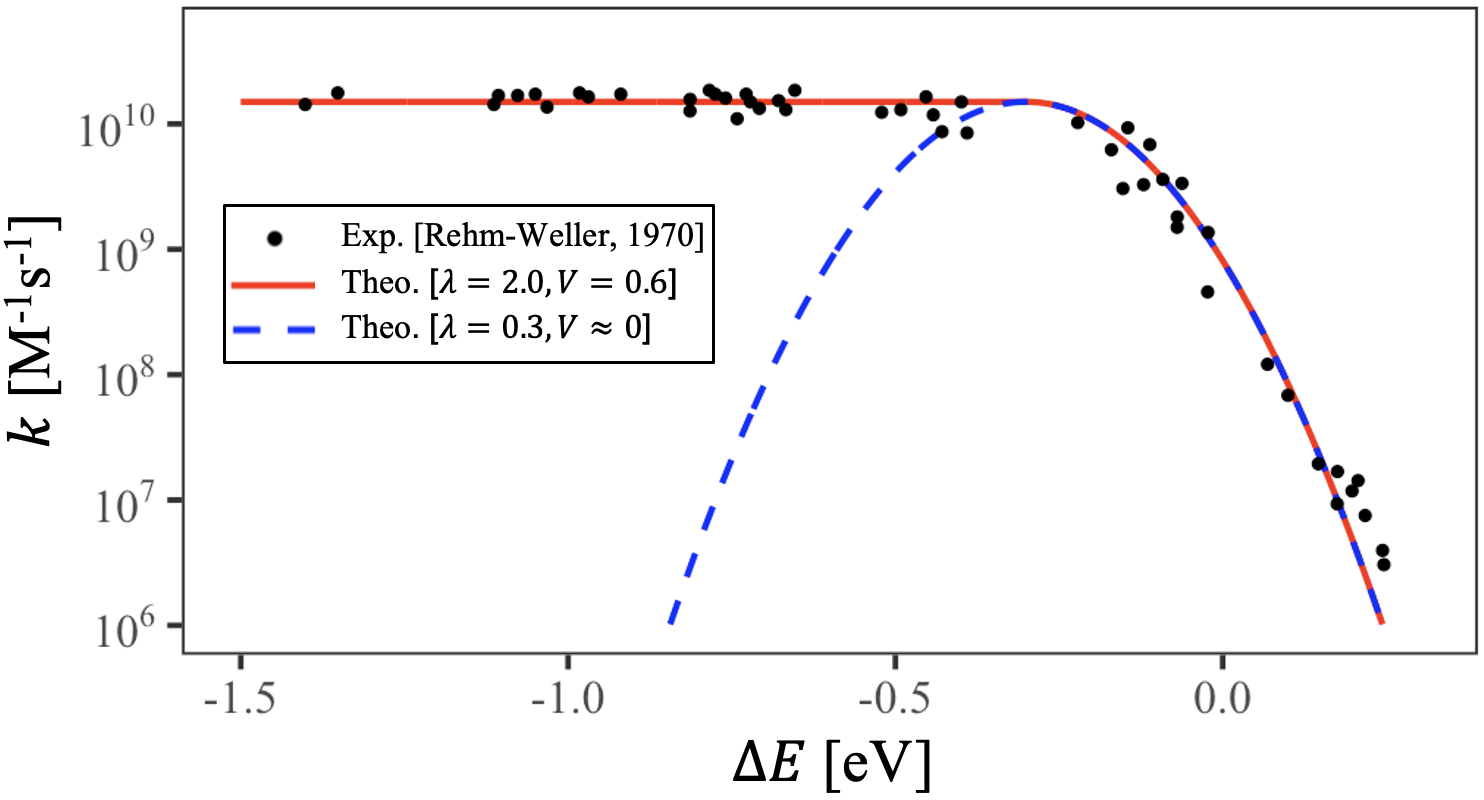}
\caption{ET rate constants $k$ as a function of the driving force $\Delta E$. The black dots are experimental data reported in Ref. \cite{experiment_Weller} (the filled circles in Fig. 2 therein) for fluorescence quenching electron transfer in acetonitrile. The red solid line shows the theoretical prediction of our two-state system with $\lambda = 2.0$ eV and $V = 0.6$ eV leading to $\Lambda=0.3$ eV (adiabatic), while the blue dashed line shows the prediction of the same system with $\lambda=0.3$ eV and $V\approx 0$ (nonadiabatic). The temperature is taken to be $298$ K for the theoretical predictions.}
\label{FIG2}
\end{figure}

Figure \ref{FIG2} shows experimental kinetic data (black dots) reported by Rehm and Weller \cite{experiment_Weller}. Using $\lambda=2.0$ eV and $V=0.6$ eV with prefactor $C=1.5\times 10^{10}$ M$^{-1}$s$^{-1}$, our model fits the Rehm–Weller kinetic data (see red solid line in Fig. \ref{FIG2}). These values of $\lambda$ and $V$ give $\Lambda=0.3$ eV via the discovered relation. Note that the values of $\lambda$ and $V$ which yield the desired $\Lambda$ are not unique; for example a similar fit can be obtained by using $\lambda=3.0$ eV and $V=1.0$ eV. Nevertheless, the conclusion that our model can explain the data via an adiabatic reaction mechanism with $\Lambda\ll\lambda$ survives the non-uniqueness of the parameter values. Repeating such experiments across solvents differing in dielectric constants, and/or across donor-acceptor pairs where something is known about the relative electronic coupling, provides a possible route to lifting this parameter degeneracy.

In contrast to this adiabatic model, we note that fitting the normal-region data with a nonadiabatic rate expression would require using $\lambda\approx0.3$ eV and $V\approx0$ with the same prefactor $C$ (see blue dashed line in Fig. \ref{FIG2}). In addition to the fact that such parameter values would predict an inverted region which is not experimentally observed, we note that this value of the diabatic reorganization energy is unusually small for electron transfer in typical polar molecular solvents, where molecular simulations generally predict substantially larger solvent reorganization energies \cite{Vladimirov_reorg,Blumberger_Sprik_2006_TCA,Qin2023_CO2_Reduction,Park_Willard_2025_PIMD_ET,acetonitrile_solvent,Parsons_reorg}. The tradeoff is that our fit requires a larger electronic coupling than is typically assumed. However, the coupling is often difficult to infer from experiment, and constrained density functional theory (cDFT) calculations \cite{Troy_cDFT_proof,Troy_mixed_cdft} have yielded larger couplings than expected for several other processes, including the carbon dioxide reduction reaction (CO$_2$RR) and charge transport in organic semiconductors \cite{Ethan_2D_landscape,Troy_disorder}.

The fact that a two-state harmonic quantum system can predict both Marcus inverted and Rehm–Weller kinetics in the nonadiabatic and adiabatic limits respectively may give new context to the prevalence of experiments in which saturation rather than inversion is observed. As Marcus recalled in his Nobel lecture, experimental evidence for the inverted-region behavior was not obtained until \textit{"almost 25 years after it was predicted"} \cite{Nobel_Marcus}. Indeed, in his original 1956 paper, Marcus explicitly stated that his theory is developed \textit{"on the basis of the assumption that there is little overlap of the electronic orbitals of the two reacting particles in the activated complex"} \cite{Marcus_1956}. Could it be that Marcus's harmonic two-state model was even more powerful than he anticipated, and that the \textit{normal-region} driving force dependence Marcus predicted holds---to a good approximation---across a broader range of coupling strengths? Could it be that the inconsistent agreement between Marcus theory and experiment in the \textit{inverted region} merely follows from the small-coupling assumption of the original treatment, whereas if the same underlying model is extended to both nonadiabatic and adiabatic regimes on equal footing, a much larger body of experimental data can be explained? That a normal-region driving force dependence similar to that which Marcus derived is predicted, within his own model, for strongly-coupled reactions may have been historically masked by the subtle mathematical transformation presented here. 

Given our proposed identification of Marcus and Rehm–Weller kinetics with the nonadiabatic and adiabatic regimes respectively, it is particularly noteworthy that inverted kinetics are seldom observed for intermolecular liquid ET \cite{experiment_Weller,experimental_Foote,experiment_Scheerer} and most often observed for long-range intramolecular ET \cite{Miller_inverted,Wasielewski_inverted,experimental_Giovanny}. It is known that electronic coupling decays approximately exponentially with the donor-acceptor distance $r$, as often modeled by the form $V(r)=V_0\text{exp}[-\beta r]$ for some constants $V_0$ and $\beta$ \cite{Troy_mixed_cdft,Marcus_1985,Ratner_review}. While it has also been shown that the diabatic reorganization energy $\lambda$ can exhibit some dependence on $r$ \cite{Oberhofer_Blumberger_2010_Angew,Sharp_1998_BiophysJ,Ethan_2D_landscape,Oberhofer_Blumberger_2010_Angew}, if one assumes that the coupling is the dominant source of the distance dependence of $\Lambda,$ we obtain

\begin{equation}\Lambda(r)=\lambda\left(1-\frac{2V_0 \text{exp}[-\beta r]}{\lambda}\right)^2.\end{equation} This result could help explain why the inverted region is typically associated with intramolecular ET where $r$ is fixed at a large value $r_\text{intra}$ such that the nonadiabatic limit $\Lambda(r_\text{intra})\approx\lambda$ is attained. In contrast, for intermolecular ET in liquids, there will often be a distance $r_\text{inter}$ yielding a coupling such that the adiabatic limit $\Lambda(r_\text{inter})\ll\lambda$ is attained, which is precisely the criterion for a Rehm–Weller plateau.

Before we conclude, there is one important subtlety worth mentioning. We note that in the rightmost panel of Fig. \ref{FIG0}(b) (i.e., corresponding to $\Delta E<-\lambda$), the mapping between diabats and adiabats has been altered, and the donor diabat now corresponds to the excited state adiabat $E_+(q)$. It has previously been noted that in this inverted region, electronic coupling will suppress rather than enhance the reaction rate \cite{adiabatic_supression,adiabatic_tunneling,Fay_extending}. Weakly adiabatic corrections in this region were also considered schematically in Ref. \cite{Bazant_CIET} (Fig. 6f). Here we add the observation that when the electronic coupling is non-perturbative (e.g., on the order of eV), the probability of a Marcus-like nonadiabatic transition in the inverted region becomes vanishingly small, and alternative mechanisms such as nuclear tunneling and radiative/non-radiative decay \cite{Englman01021970,kramer_review, Nitzan_2006}, or pathways to alternative reaction products, may dominate the rate. 

To make the origin of this change in the inverted mechanism due to the topological alteration of the mapping between adiabats and diabats more concrete, we recall that when the temperature is finite, the probability of the system leaving the higher energy adiabat when thermally excited to the avoided crossing can be approximated by the Landau-Zener formula \cite{Nitzan_2006,Zener1932,Landau_theory}. The probability of such a nonadiabatic transition at a crossing event is given by

\begin{equation}\label{inverted:13}
P_\text{NA} =\text{exp}\left(-\frac{2\pi V^{2}}{\hbar \dot{q}\, |\Delta F|}\right),
\end{equation}
where $\dot{q}=\partial q/\partial t$ is drawn from a temperature-dependent Boltzmann distribution and $\Delta F=(\partial E_A/\partial q-\partial E_D/\partial q)|_{q=q^*}.$ Indeed, the latter can easily be evaluated from $E_D(q)$ and $E_A(q)$ given in Eq. (\ref{energies}). Interestingly, we obtain \begin{equation}\label{inverted:14}
|\Delta F|= 2\lambda,\end{equation} which shows how the probability of a nonadiabatic transition decays exponentially with the second order mixing parameter $V^2/\lambda$. That is, \begin{equation}\label{inverted:15}P_\text{NA}\approx\text{exp}[-\alpha (V^2/\lambda)],~~\text{with}~ \alpha=\pi/(\hbar \dot{q}).\end{equation} The standard reaction rate in this inverted region would therefore be approximated by \begin{equation}\label{inverted:16}k_\text{ET} \approx \nu \exp\!\left\{ -\frac{\alpha V^{2}}{\lambda} - \frac{(\lambda + \Delta E)^{2}}{4\lambda k_{\mathrm{B}}T} \right\},\end{equation} where $\nu$  is a classical attempt frequency. The adiabatic limit implies $\alpha V^2/\lambda\gg1$, and so $k_\text{ET}\rightarrow0$ regardless of the value of the Marcus activation factor. This simple analysis supports the idea that in this adiabatic inverted region, the reaction may occur via an entirely alternative mechanism, possibly orders of magnitude slower than the Marcus mechanism. We note that the prediction that has emerged from our analysis, of three regions with distinct behavior as the driving force is increased---(1) Marcus normal kinetics, (2) extended plateau, (3) rapid decrease in rate---has already been observed experimentally in Ref. \cite{Three_Regions}.

In the opposite, nonadiabatic limit $V \to 0$, the plateau window narrows as
$\lambda - \Lambda = 4V - 4V^2/\lambda \approx 4V$ and vanishes, and the system
returns to the regime in which the standard nonadiabatic Marcus rate, with its
familiar inverted-region dependence on driving force, applies. The Marcus
inverted region is therefore recovered continuously as the coupling decreases,
consistent with its observation in weakly coupled, long-range
systems~\cite{Miller_inverted,Wasielewski_inverted,experimental_Giovanny,experimental_Sun}.

In conclusion, we have extended the Marcus model to adiabatic reactions and found that in this limit, Rehm–Weller kinetics are predicted mathematically. We have also demonstrated that a parameter $\Lambda$---hidden in the mathematics of the familiar model---can extend the Marcus-like driving-force dependence to the adiabatic limit, consistent with the numerical examples in Ref. \cite{Ethan_lambda_eff}. To what extent the mathematical prediction presented here contributes to the Rehm–Weller effect, as compared to the many other explanations that have been proposed, is a subject ripe for further investigation. In particular, separating the parameters $\lambda$ and $V$ using experimental and computational approaches will provide crucial insight into the behavior. Similarly, experimental and theoretical analysis of the adiabatic inverted region is an open research topic \cite{Fay_extending}. Tunable Hamiltonian model systems, where both the driving force and electronic coupling can be systematically varied \cite{Marcus_Organic_Semiconductor,Experimental_Holstein_1,Experimental_Holstein_2,Marcus_Organic_Semiconductor}, may provide a route toward direct testing of the presented theory. 

Beyond the relevance for Rehm–Weller kinetics considered here, a series of forthcoming studies by the author and coworkers will also show that the effective reorganization energy $\Lambda$ for electrochemical systems can be computed from constrained quantum simulations \cite{Ethan_2D_landscape,Troy_cDFT_proof,Troy_mixed_cdft}, and that extending an earlier weak-coupling theory of electrochemical kinetics \cite{Bazant_CIET}, it is $\Lambda$, rather than the classical $\lambda$, that governs the Marcus-like current-voltage response of industrial Faradaic reactions such as carbon dioxide reduction and battery intercalation.

\section*{Acknowledgments} The author is grateful to Martin Bazant, Abraham Nitzan, Troy Van Voorhis, and Junghyun Yoon for useful discussions.

\section*{Author Declaration} The author has no conflicts of interest to declare.

\end{document}